\documentclass[twocolumn]{aastex631}
\usepackage{amsmath}
\usepackage{subfigure}
\usepackage{xcolor}
\usepackage{float,color}
\usepackage{gensymb}

\usepackage{mdframed}

\begin{document}

\title{Rates and beaming angles of GRBs associated with compact binary coalescences}

\author[0000-0001-5318-1253]{Shasvath J. Kapadia}
\affiliation{Inter-University Centre for Astronomy and Astrophysics, Post Bag 4, Ganeshkhind, Pune - 411007, India}

\author[0000-0001-9868-9042]{Dimple}
\affiliation{Chennai Mathematical Institute, Siruseri, 603103 Tamilnadu, India}
\affiliation{Institute for Gravitational Wave Astronomy and School of Physics and Astronomy, University of Birmingham, Birmingham, B15 2TT,}

\author{Dhruv Jain}
\affiliation{Aryabhatta Research Institute of Observational Sciences (ARIES), Manora Peak, Nainital-263002, India}

\author[0000-0003-1637-267X]{Kuntal Misra}
\affiliation{Aryabhatta Research Institute of Observational Sciences (ARIES), Manora Peak, Nainital-263002, India}

\author[0000-0002-6960-8538]{K. G. Arun}
\affiliation{Chennai Mathematical Institute, Siruseri, 603103 Tamilnadu, India}

\author{L. Resmi}
\affiliation{Indian Institute of Space Science $\&$ Technology, Trivandrum 695547, India}


\begin{abstract}
 Some, if not all, binary neutron star (BNS) coalescences, and a fraction of neutron - star black hole (NSBH) mergers, are thought to produce sufficient mass-ejection to power Gamma-Ray Bursts (GRBs). However, this fraction, as well as the distribution of beaming angles of BNS-associated GRBs, are poorly constrained from observation. Recent work applied machine learning tools to analyze GRB light curves observed by {\textit{Fermi}}/GBM and {\it Swift}/BAT. GRBs were segregated into multiple distinct clusters, with the tantalizing possibility that one of them (BNS cluster) could be associated with BNSs and another (NSBH cluster) with NSBHs. As a proof of principle, assuming that all GRBs detected by {\it Fermi}/GBM and {\it Swift}/BAT associated with BNSs (NSBHs) lie in the BNS (NSBH) cluster, we estimate their rates ($\mathrm{Gpc}^{-3}\mathrm{yr}^{-1}$). We compare these rates with corresponding BNS and NSBH rates estimated by the LIGO-Virgo-Kagra (LVK) collaboration from the first three observing runs (O1, O2, O3). We find that the BNS rates are consistent with LVK's rate estimates, assuming a uniform distribution of beaming fractions ($f_b \in [0.01, 0.1]$). Conversely, using the LVK's BNS rate estimates, assuming all BNS mergers produce GRBs, we are able to constrain the beaming angle distribution to $\theta_j \in [0.8^{\circ}, 33.5^{\circ}]$ at $90\%$ confidence. We similarly place limits on the fraction of GRB-Bright NSBHs as $f_B \in [1.3\%, 63\%]$ ($f_B \in [0.4\%, 15\%]$) with {\it Fermi}/GBM ({\it Swift}/BAT) data.

\end{abstract}



\section{Introduction} \label{sec:intro}

The LIGO-Virgo-Kagra detector network \citep{LIGODetector, VirgoDetector, KAGRADetector} has observed $\sim 90$ compact binary coalescences (CBCs) during its first three observing runs (O1, O2, O3) \citep{KAGRA:2021vkt}. Among these, only two were binary neutron star (BNS) mergers \citep{LIGOScientific:2017vwq, LIGOScientific:2020aai}, of which only one (GW170817) produced an observed kilonova and Gamma-Ray Burst (GRB) \citep{LIGOScientific:2017ync}. Two neutron star - black hole mergers (NSBHs) were also detected, but no corresponding EM counterparts were observed \citep{LIGOScientific:2021qlt}. 

The prevalent expectation is that some, if not all, BNS mergers produce EM counterparts, including GRBs \citep{Paczynski_1986, Narayan_1991, embright}. On the other hand, not all NSBH mergers are expected to produce such counterparts \citep[see, e.g.][and references therein]{foucart-nsbh}. A necessary (though not sufficient) condition for producing these counterparts is a non-zero remnant mass post-merger \citep{foucart2012}. Intuitively, if the tidal forces on the NS from the BH are sufficiently large, then the NS can disrupt outside the innermost stable circular orbit (ISCO) of the BH, which in turn will leave a non-zero remnant mass. 

Such a requirement imposes a restriction on the mass-ratios $q$ of the NSBH that could allow such an extra-ISCO NS disruption, although the exact $q$ values ultimately depend on the equation of state of the NS, and the spin of the BH \citep{foucart2018}. Nevertheless, lower mass ratios (with the mass ratio defined such that $q \in [1, \infty)$) are more likely to produce EM-counterparts \citep{foucart2013}, including GRBs, though their occurrence is not guaranteed given the complexities associated with the physics of jet production \citep[see, e.g.][]{salafia2020gamma}. 

Testing standard assumptions about GRB-Bright CBCs ideally requires multiple confident joint GW-GRB detections, of which there is a severe paucity \citep{LIGOScientific:2021iyk, LIGOScientific:2020lst, LIGOScientific:2019obb, FermiGamma-rayBurstMonitorTeam:2018eao}. However, a number of GRBs have been detected by the gamma-ray burst monitor (GBM) on board the \textit{Fermi} satellite, as well as the Burst Alert Telescope (BAT) of the \textit{Swift} satellite. GRBs are typically classified into long and short GRBs, based on the duration $T_{90}$ for $90\%$ of the energy in prompt emission to be released. The provenance of long GRBs (lGRBs, $T_{90} > 2s$) has traditionally been attributed to collapsars (collapse of massive stars) \citep{Woosley1993, Paczynski1998, Mezaros2006, Woosley2006, hjorth2012grb}, while short GRBs (sGRBs, $T_{90} < 2s$) are thought to be produced by CBCs \citep{Paczynski1986, Eichler1989, Meszaros1992}. These assumptions have been supported by observations of broad-lined Type Ic supernovae (SNe) coincident with lGRBs \citep{Galama1998, Kulkarni1998, Stanek2003, Cano2017}, and r-process nucleosynthesis driven Kilonovae (KNe) associated with sGRBs \citep{Berger2013, Tanvir2013, Kasliwal2017, Abbott2017, Valenti2017, Troja2019, Lamb2019}.

However, some GRB events were found to violate standard expectations. Examples include GRB~200826A, which was a sGRB with an SN counterpart \citep{Ahumada2021, Zhang2021}, GRBs~060614 and 060505 were lGRBs with no observed SNe associated with them \footnote{The lack of SNe is not entirely unexpected for high-redshift GRBs.} \citep{Fynbo2006}, and GRBs 060614 
\citep{Yang2015} and 211227A \citep{Lu2022} which were long but data suggested evidence of coincident KNe. Indeed, even data from the observations of high-redshift sGRBs suggest collapsars as potential progenitors of these events \citep{Dimple2022}. 

Recently, GRB~211211A had a prolonged burst phase with an associated KN \citep{Rastinejad2022}. Several models have been proposed to suggest its provenance, most of which point to a CBC-origin \citep{Yang2022, Gompertz:2022jsg, Troja2022, Zhu2022, Zhong2023, Kunert:2023vqd}. This conclusion is further corroborated by the host galaxy whose properties coincided with those pertaining to hosts of sGRBs \citep{Troja2022, Yang2022}. Moreover, the GRB's location with respect to the host's center seemed to suggest a lack of spatial coincidence with any star forming region, which again points to a CBC progenitor rather than a collapsar origin \citep{Fruchter2006, Berger2013}. More recently, JWST observations of the bright lGRB~230307a also indicated the presence of a potential KN \citep{Sun_2023, Levan_2024}.

It is, therefore, amply evident that the standard picture of sGRB + KN + CBC and lGRB + SN + collapsar cannot be reconciled with data that increasingly suggests a more complex picture. A more sophisticated approach to segregating GRBs into groups and subgroups belonging to a collapsar origin or a CBC origin and to further split the CBC group into a BNS or NSBH cluster is to use machine learning (ML). Several works have attempted this, where light curves from the prompt emission phase of GRBs detected by one instrument are fed to an unsupervised ML algorithm that identifies global structures \citep{Jespersen2020, Steinhardt:2023jhm, Luo:2022gur, Garcia-Cifuentes:2023jko}. 
Recently, \citet{Dimple:2023wvs} identified five distinct clusters in the {\it Swift}/BAT GRB population, with KN-associated GRBs located in two separate clusters. This structure remains true for both the catalogs -- {\it Swift}/BAT as well as for {\it Fermi}/GBM \citep{Dimple_2024}. They suggested that these two distinct populations might be associated with BNS and/or NSBH mergers. One cluster with KN-associated GRBs having longer $T_{90}$'s might be associated with NSBH mergers 
\citep{Zhu2022}. On the other hand, the presence of GRB~170817A in the other cluster \citep{Dimple_2024} confirms that at least a fraction of GRBs in that cluster might be associated with BNS mergers.

While there is some uncertainty on whether all GRBs that sit in either of these two clusters can be associated with a BNS or NSBH, it is still interesting to estimate the rate of BNS/NSBH- associated GRBs, assuming that such GRBs must lie in these two clusters. In this {\it Letter}, we evaluate these rates, and compare them to the BNS and NSBH rates reported by the LVK in its first three observing runs (O1, O2, O3) \citep{KAGRA:2021duu}. We find that the rate of BNS-associated GRBs are broadly consistent with the BNS-associated GW rate estimates. We additionally constrain the fraction of GRB-bright NSBHs, and, assuming that all BNSs produce GRBs, the beaming angle of BNS-associated GRBs. 

\section{Method}
\subsection{Estimating the Rate of GRBs}
Most works that estimate the rate of GRBs assume a model for the redshift evolution, which is typically chosen to be the Madau-Dickinson ansatz \citep{Madau:2014bja, madau2017radiation} pertaining to star formation rate. While this model may be sufficient for GRBs thought to originate from collapsars, a model for the rate of CBC-driven GRBs needs to additionally account for a delay time between the formation of the compact binary and its coalescence \citep{wanderman2015rate}. This is usually achieved by adopting an additional ansatz on the distribution of delay times, although there are considerable uncertainties in the models of delay-time distributions.

On the other hand, the rate of BNS and NSBH mergers estimated by the LVK ($R_{\rm GW}$), given GW data ($d_{\rm GW}$), is limited by the relative scarcity of BNS/NSBH-associated GW detections \citep{KAGRA:2021vkt}. Moreover, the confirmed BNS/NSBH-associated GW detections are at relatively low redshifts, making it difficult to probe their redshift evolution\footnote{The GW rates also incorporate subthreshold BNS and NSBH events that might have arrived from larger distances, but their low signal-to-noise ratios make it difficult to constrain their luminosity distances, and therefore redshifts.}. We therefore similarly evaluate a rate of BNS/NSBH-associated GRBs ($R_{\rm GRB}$) assuming that the sources are distributed uniformly in comoving volume, as done in Appendix C of \cite{KAGRA:2021duu}, \cite{LIGOScientific:2021qlt}, and \cite{Kapadia:2019uut}. This will allow for a self-consistent evaluation of the fraction of GRB-Bright NSBHs, as well as the beaming angle of BNS-associated GRBs.

Consider $N_{\rm GRB}$ GRB detections collected over an observation time $T_{\rm obs}$. We make two assumptions. The first is that the GRB detections are independent and, therefore, follow a Poisson process. The second is that the intrinsic luminosity function of the GRBs, $p_{L}(L)$, has no redshift dependence. 

The first assumption allows us to write down the Poisson likelihood as (we remove the GRB subscript to reduce notational burden): 
\begin{equation}
p (N | \Lambda) = \frac{1}{N!}\Lambda^N\exp(-\Lambda).
\end{equation}
The posterior on the Poisson mean is found using Bayes' rule: $p(\Lambda | N) \propto p(\Lambda)p(N | \Lambda)$. The posterior on the rate of GRBs, $R = \frac{\Lambda}{\langle VT \rangle}$, is then:
\begin{equation}
p(R | N) \propto p(\Lambda | N) = p(R\cdot \langle VT \rangle | N)
\end{equation}
The task now is to estimate the spacetime volume sensitivity, $\langle VT \rangle$ of the GRB detector, averaged over the expected distribution of fluxes $p_{F}(F)$ accessible to the detector. This sensitivity is given by: 
\begin{equation}\label{Eq:VT}
\langle VT \rangle = T_{\rm obs}\cdot\int_{0}^{\infty} \frac{1}{1 + z}\frac{dV_c}{dz}f(z) dz
\end{equation}
where $\frac{dV_c}{dz}$ is the differential comoving volume, $\frac{1}{1 + z}$ accounts for cosmological time dilation, and $f(z)$ is the fraction of spacetime volume accessible to the detector (with threshold flux $F_t$) given by:
\begin{equation}
f(z) = \int_{F_t}^{\infty} p_F(F)dF
\end{equation}
By definition, the flux $F$ and the intrinsic luminosity $L$ of the source are related by:
\begin{equation}
F = \frac{L}{4\pi d_L^2}
\end{equation}
where $d_L$ is the luminosity distance. The flux function can then be written in terms of the intrinsic luminosity function ($p_L(L)$), and $f(z)$ is evaluated as:
\begin{equation}
f(z) = 4\pi d_L^2(z) \int_{F_t}^{\infty} p_L(4\pi d_L^2(z)F) dF
\end{equation}
Let $f_{b}$ be the beaming fraction, $f_{s}$ be the fraction of the sky viewed by the detector at any given time, and $f_{k}$ the K-correction factor due to redshifting and finite bandwidth of the {\it Fermi} detector. The spacetime volume sensitivity accessible to the instrument is then Eq.~(\ref{Eq:VT}) multiplied by these fractions. Since the beaming fractions, especially for sGRBs, are poorly constrained from observation \citep{Fong2015}, we draw them from a broad, uninformative prior distribution. 

To estimate the beaming fraction of BNS-associated GRBs, we assume that {\it all} BNS mergers produce GRBs. This may be justified by the high extent of tidal disruption possible in BNS mergers compared to, say, NSBH mergers. The disruption provides enough matter near the merger site to produce an accreting central engine that can power the jet. 

We then evaluate the ratio distribution  $p(\langle f_b \rangle \equiv \frac{R_{\rm GRB}}{R_{\rm GW}} | N_{\rm GRB}, d_{\rm GW})$, pertaining to BNSs. This distribution is constructed from the GRB ($p(R_{\rm GRB} | N_{\rm GRB})$) and GW ($p(R_{\rm GW} | d_{\rm GW})$) rate posteriors. We ensure that $R_{\rm GRB}$ is evaluated with $\langle VT \rangle$ using Eq.~(\ref{Eq:VT}), {\it without} multiplying it by an assumed $f_b$. This ratio distribution constrains the average beaming fraction $\langle f_b \rangle$. 

From the ratio distribution, we evaluate constraints on the minimum ($f_b^{\mathrm{min}}$) and maximum ($f_b^{\mathrm{max}}$) beaming fractions, assuming a uniform distribution $f_{b} \in \mathcal{U}(f_b^{\mathrm{min}}, f_b^{\mathrm{max}})$. We do so using Monte Carlo, and by exploiting the fact that $\langle f_b \rangle = (f_b^{\mathrm{min}} + f_b^{\mathrm{max}})/2$. We draw from $p(\langle f_b \rangle | N_{\rm GRB}, d_{\rm GW})$. For each $\langle f_b \rangle$ sample, we draw an $f_b^{\mathrm{min}}$ sample from a uniform distribution $\mathcal{U}(0, \langle f_b \rangle)$. We then acquire an $f_b^{\mathrm{max}}$ sample as $f_b^{\mathrm{max}} = 2\langle f_b \rangle - f_b^{\mathrm{min}}$. Repeating this process iteratively, we construct distributions for $f_b^{\mathrm{min}}$ and $f_b^{\mathrm{max}}$. 

The beaming angle ($\theta_j$) is acquired from the beaming fraction via:
\begin{equation}
f_b = 1 - \cos(\theta_j)
\end{equation}
This equation assumes a simple homogeneous tophat jet structure \footnote{The tophat model for the jet is a simplifying assumption. For example, the sGRB associated with GW170817 \citep{LIGOScientific:2017zic}, exhibited a departure from this model.} The constraint on the fraction of GRB-Bright NSBHs, $f_B$ is similarly acquired from a ratio distribution pertaining to NSBHs, but now assuming a prior distribution on $f_b$.

\begin{figure*}
    \begin{mdframed}
        \centering
        \includegraphics[width=\linewidth,trim={2cm 2cm 2cm 2cm},clip]{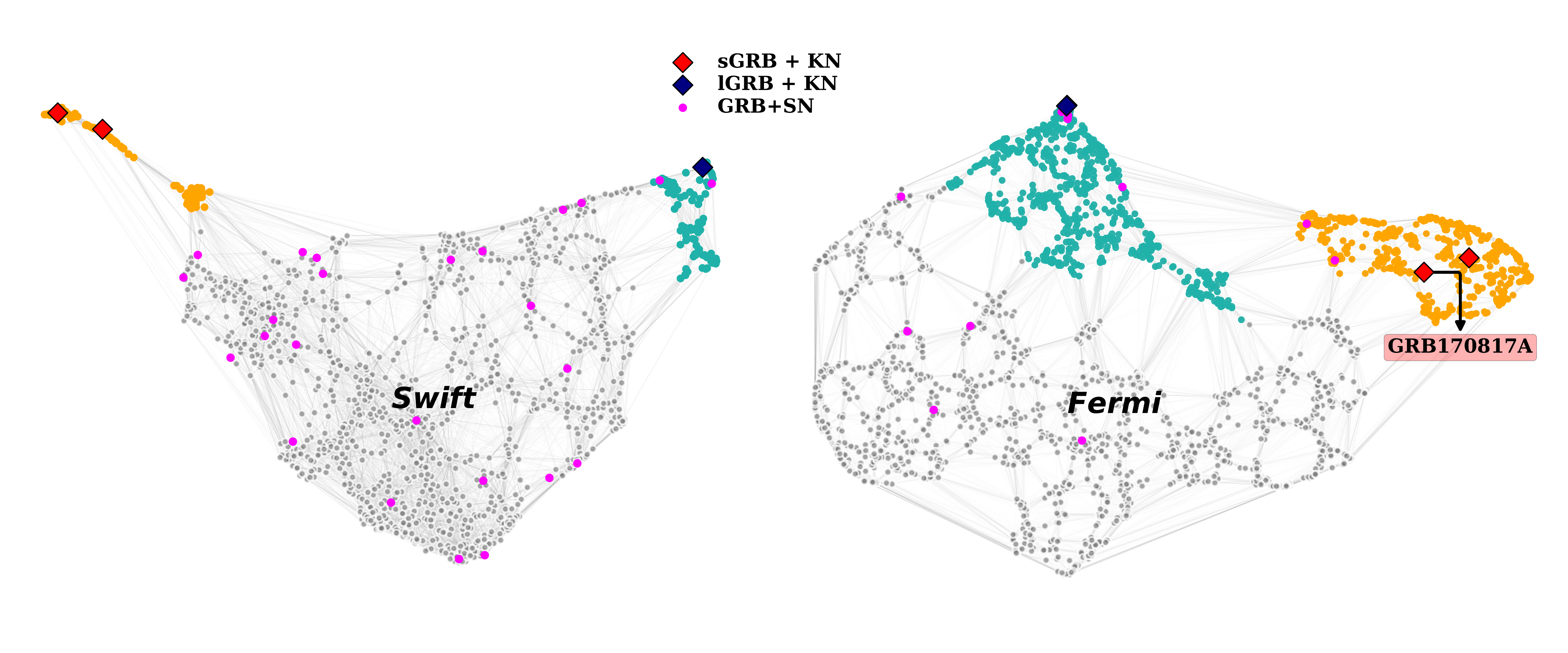}
     \end{mdframed}
    \caption{Two-dimensional embeddings obtained using PCA-UMAP for \textit{Swift}/BAT and \textit{Fermi}/GBM catalogs. The red squares show the locations of KN-associated lGRBs. The blue squares show the locations of KN-associated sGRBs. GRB~170817A is highlighted in the red label, suggesting the origin of neighboring GRBs to be BNS mergers. Therefore, we assume clusters highlighted in orange to be a BNS cluster, while those in cyan to be an NSBH cluster. The SN-associated GRBs are represented by the magenta color circles. Notice that some of these clusters contain known false-positives. (See \citealt{Dimple_2024, Dimple:2023wvs} for details.)} 
    \label{fig:ML_results}
\end{figure*}

\subsection{Progenitor Identification with Machine Learning}
To estimate BNS/NSBH-associated GRB rates, it is crucial to first identify, from the set of GRBs detected by an instrument, those that can be associated with a BNS or an NSBH. \citet{Dimple:2023wvs} used machine learning algorithms, particularly uniform manifold approximation and projection \citep[UMAP,][]{McInnes2018}, and t-distributed stochastic neighbor embedding  \citep{maaten2008visualizing, van2014accelerating}, with a principle component analysis (PCA) initialization on GRB light curves in \textit{Swift}/BAT, to achieve this. Although working on different principles, both algorithms separated the KN-associated GRBs into two distinct clusters. This suggested two distinct populations of KN-associated GRBs, one associated with sGRBs and the other with lGRBs. Further inspection suggested they might have different origins (see Figure 4 in \citealt{Dimple:2023wvs}).

The association of clusters with a BNS or NSBH provenance becomes stronger with data from the \textit{Fermi}/GBM catalog \citep{Dimple_2024}. Figure \ref{fig:ML_results} shows the 2-dimensional embeddings obtained using PCA-UMAP on \textit{Swift}/BAT and \textit{Fermi}/GBM light curves. The one KN-associated sGRB event, detected by {\it Fermi}, whose BNS provenance is beyond all ``reasonable'' doubt, GW170817's counterpart, GRB~170817A, lies within what we call the ``BNS cluster'' in the embedding. This implies that some fraction of GRBs in that cluster might originate from BNS mergers. For GRBs in the other ``NSBH cluster'', some have coincident KNe and evidence of fallback accretion. This suggests that NSBH mergers could be progenitors \citep{Zhu2022} for at least a fraction of the GRBs in this cluster. However, only multimessenger observations of NSBH coalescences can confirm this prediction.

While not confirmed, we nevertheless still work with the assumption that all BNS/NSBH- associated GRBs in the data set (from {\it Fermi} or {\it Swift}) have been identified and placed in the BNS/NSBH clusters. However, not all events in these clusters have a BNS/NSBH provenance. Some known SNe also lie in these clusters. We account for this when estimating BNS/NSBH-associated GRB rates, as delineated in the following subsection.


%
\begin{figure*}[htb]
    \includegraphics[width=0.5\linewidth]{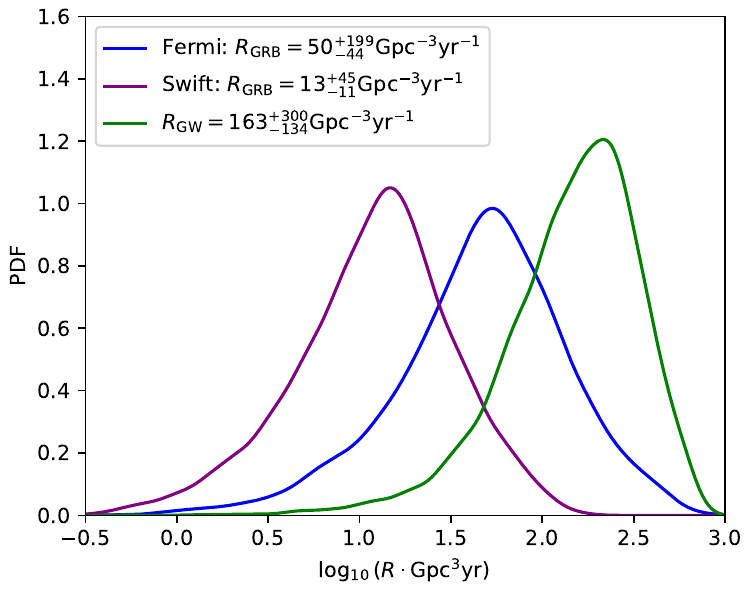}
    \includegraphics[width=0.5\linewidth]{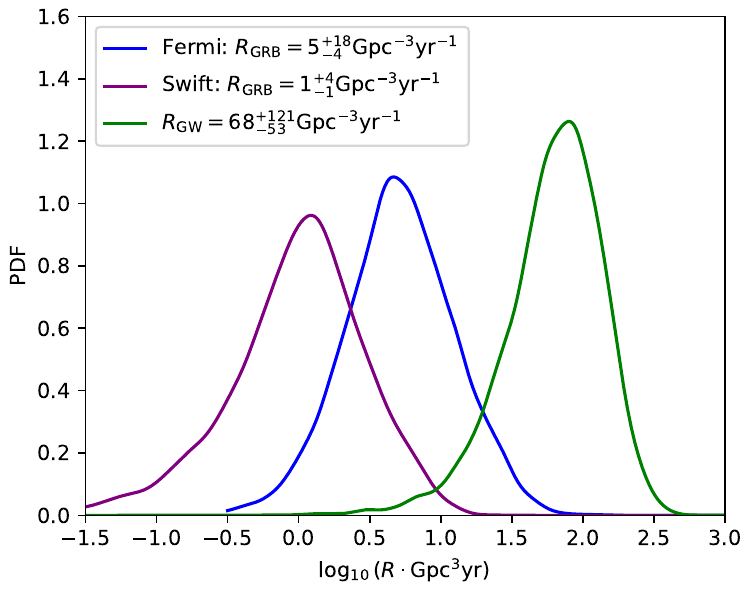}
    \caption{{\it Left Panel:} The posteriors on the rate of GRB-Bright BNS mergers, $p(R_{\mathrm{GRB}} | \mathcal{N}_{\mathrm{BNS}})$, estimated using {\it Fermi} data and {\it Swift} data. These are both broadly consistent with GW BNS rate posterior $p(R_{\mathrm{GW}} | d_{\mathrm{GW}})$. The $90\%$ rate estimates are quoted on the top left corner of the panel. The larger errorbar on the {\it Fermi} rate estimates can be attributed to the larger uncertainty in the fraction of events in the BNS cluster that have a BNS provenance, as compared to the {\it Swift} rate estimates. {\it Right Panel:} Same as the {\it Left Panel}, but for NSBHs. The {\it Fermi} and {\it Swift} rate estimates are broadly consistent with each other, but are markedly smaller than the GW NSBH rate posterior, as expected.}
    \label{fig:rates}
\end{figure*}

\subsection{Counting Error Estimation}


Let $n$ be the total number of events in the BNS or NSBH cluster. Of these, let $N_{\mathrm{KN}}$ be the number of known KNe and let $N_{\mathrm{SN}}$ be the number of known SNe. We wish to evaluate the probability that there are $N$ KNe in the cluster, given that $N$ could lie anywhere between $N_{\mathrm{KN}} \leq N \leq n - N_{\mathrm{SN}}$:
\begin{equation}
p(N | \mathcal{N}), ~~ \mathcal{N} \equiv N_{\mathrm{KN}} \leq N \leq n - N_{\mathrm{SN}}
\end{equation}
From Bayes' theorem:
\begin{equation}\label{Eq:Bayes}
p(N | \mathcal{N}) = \frac{p(\mathcal{N} | N)p(N)}{p(\mathcal{N})}
\end{equation}
The prior $p(N)$ is assumed to be uniform for $N$ satisfying $\mathcal{N}$, and zero otherwise. The likelihood can be written as:
\begin{equation}\label{Eq:likelihood}
p(\mathcal{N} | N) = \int_0^{1}p(\mathcal{N}|q)p(q | N)dq
\end{equation}
where $q$ is the probability that an individual event is a KN, which is unknown. From Bayes' theorem:
\begin{equation}
p(q | N) \propto p(N|q)p(q)
\end{equation}
Assuming $p(N | q)$ is Binomially distributed, and a uniform prior on $q$, we get, after normalisation:
\begin{equation}\label{Eq:prob_succ_trial}
p(q | N) = \frac{q^N(1 - q)^{n - N}}{\beta(N + 1, n - N + 1)}
\end{equation}
where the function in the denominator is the Euler Beta function.
%
%

On the other hand, it is straightforward to see that:
\begin{equation}\label{Eq:binomial}
p(\mathcal{N} | q) = 1 - \sum_{j = 0}^{N_{\mathrm{KN}} - 1} B(j; n, q) - \sum_{j = 0}^{N_{\mathrm{SN}} - 1} B(n - j; n, q)
\end{equation}
where $B(j;n,q)$ is the usual Binomial distribution with $j$ successful trials. Thus, $p(N | \mathcal{N})$ can be evaluated from Eqs.~\ref{Eq:Bayes}, \ref{Eq:likelihood},  \ref{Eq:prob_succ_trial}, and \ref{Eq:binomial}. 

The posterior on the Poisson mean is then convolved with $p(N|\mathcal{N})$, thereby incorporating counting uncertainties:
\begin{equation}
p(\Lambda | \mathcal{N}) = \sum_{N = 0}^{n}p(\Lambda | N)p(N | \mathcal{N})
\end{equation}
%
 
\section{Results and Discussion}

\begin{figure*}[htb]
    \includegraphics[width=0.5\linewidth]{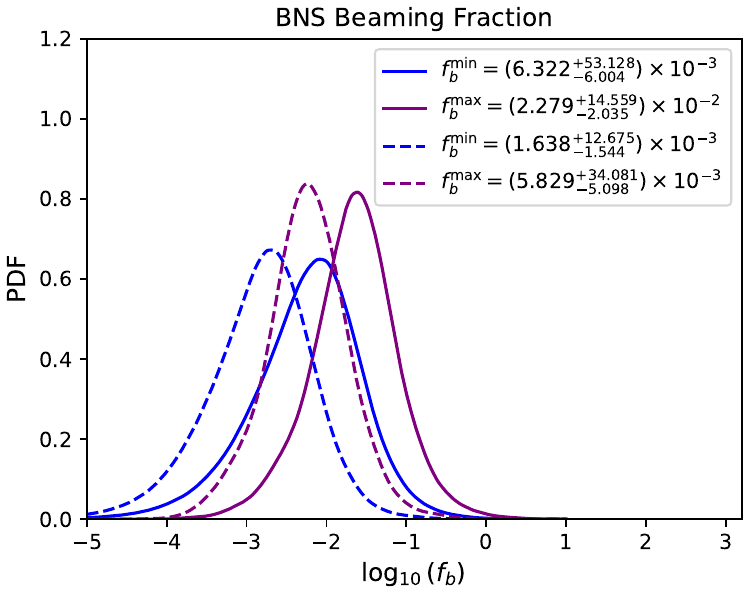}
    \includegraphics[width=0.5\linewidth]{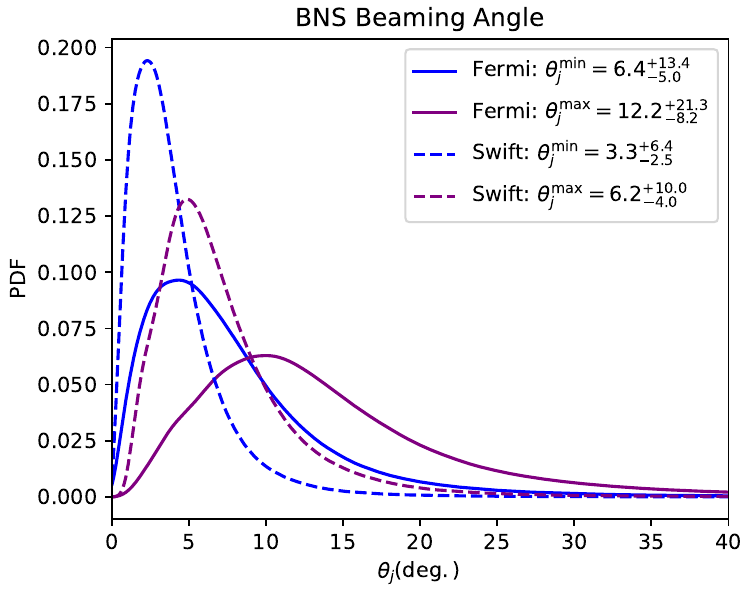}
    \caption{{\it Left Panel}: Constraints on the distribution of beaming fractions for GRBs with a BNS provenance, assuming that all BNSs produce GRBs. The constraints from {\it Fermi} are less stringent than those from {\it Swift}. The reason is the same as what was described in Figure~\ref{fig:rates} to explain the broader rate posteriors for {\it Fermi} as compared to {\it Swift}. {\it Right Panel}: Distributions of beaming fractions in the {\it Left Panel} converted to distributions of beaming angles $\theta_j$ (in $\degree$). These span $\theta_j \in [0.8 \degree, 33.5 \degree]$ across both {\it Fermi} and {\it Swift} estimates.}
    \label{fig:fraction}
\end{figure*}

We present the results of our estimates of the rate of BNS/NSBH-associated GRBs. We use {\it Fermi} data \citep{von2020fourth}, with observing time $T_{\rm obs} = 15.22 ~{\rm yrs}$, and {\it Swift} data \citep{lien2016third} with observing time $T_{\rm obs} = 18.8 ~{\rm yrs}$. The ML classification scheme of \citet{Dimple_2024} identifies sGRB and lGRB events in the BNS and NSBH clusters, which we tabulate in Table 1. We also point out the number of known KNe and SNe in those clusters, in Table 1., as ``(no. known KNe/ no. known SNe)''.

\begin{table}[ht]\label{Tab:Data}
\centering
\begin{tabular}{|c|c|c|c|c|}
\hline
 & \multicolumn{2}{c|}{\textbf{\textit{Fermi}}} & \multicolumn{2}{c|}{\textbf{\textit{Swift}}} \\ \hline
 & BNS & NSBH & BNS & NSBH \\ \hline
\textbf{lGRB} & 141(0/1) & 707(2/7) & 20(0/0) & 115(1/4) \\ \hline
\textbf{sGRB} & 452(2/1) & 9(0/0) & 97(2/0) & 0(0/0) \\ \hline
\end{tabular}
\caption{Number of sGRBs and lGRBs, in by \textit{Fermi} and \textit{Swift} data, identified by the ML classification scheme of \cite{Dimple_2024} in the BNS and NSBH clusters. The (no. known KNe/ no. known SNe) in each of these clusters is also tabulated.}
\end{table}

We use these numbers to estimate the posterior on the Poisson mean $\Lambda$, $p(\Lambda | \mathcal{N})$, as described in the previous section. We then convert this posterior to one on the rate by evaluating the spacetime volume sensitivity $\langle VT \rangle$. We use Eq.~\ref{Eq:VT} to estimate $\langle VT \rangle$, assuming a flux threshold $F_t$ of $2 \times 10^{-7} \mathrm{ergs}/\mathrm{s}/\mathrm{cm}^2$ ($5 \times 10^{-8} \mathrm{ergs}/\mathrm{s}/\mathrm{cm}^2$) for {\it Fermi}/GBM ({\it Swift}/BAT).

sGRBs and lGRBs have different luminosity functions. We therefore calculate $\langle VT \rangle$, and the rate of BNS/NSBH associated sGRBs and lGRBs, separately, using their respective luminosity functions, and numbers in each of the clusters tabulated in Table 1. We then add the two rates to quote a total BNS/NSBH associated GRB rate. We assume a broken power-law profile for the luminosity function $p_L(L)$, as done in \cite{wanderman2010luminosity, wanderman2015rate}:
\begin{equation}
     \phi(L)=\begin{cases}
    \left(\frac{L}{L_b}\right)^{-a}, & \text{if $L \leq L_b$}.\\
    \left(\frac{L}{L_b}\right)^{-b}, & \text{if $L > L_b$}.
  \end{cases}
\end{equation}
where $\phi(L)$ traditionally denotes the number of samples between $\log_{10}(L)$ and $\log_{10}(L) + d\log_{10}(L)$ \citep{wanderman2010luminosity}, and can be straightforwardly related to $p_L(L)$. The shape parameters of the broken power-law function, for lGRBs (sGRBs), estimated by \cite{wanderman2010luminosity} (\cite{wanderman2015rate}) are $a = 0.95^{+ 0.12}_{- 0.12}$ ($0.2^{+0.2}_{-0.1}$), $b = 2.0^{+ 1.0}_{- 0.8}$ ($1.4^{+0.3}_{-0.6}$), and $L_b = 2.0^{+1.4}_{-0.4} \times 10^{52}$ ($10^{52.5 \pm 0.2}$) ergs/s.

We construct distributions on $\langle VT \rangle$ by approximating the errors on the shape parameters as Gaussians, and drawing from these Gaussians assuming that the values provided pertain to medians and upper/lower limits at $1\sigma$ confidence\footnote{When the upper/lower limits are asymmetric, we take the larger of the two, in absolute value, as the standard deviation of the Gaussian.}. We also consider a uniform distribution of $f_b \in [0.01, 0.10]$. Moreover, we take the K-correction factor as $f_K = 1.13$ \citep{Poolakkil_2021} and sky fraction to be $f_s = 2/3$ ($f_s = 1/6$) for {\it Fermi}/GBM ({\it Swift}/BAT).

We draw $10^4$ samples from $p(\Lambda | \mathcal{N})$ and the $\langle VT \rangle$ distribution, and divide sample by sample to estimate the GRB rate posterior $p(R_{\mathrm{GRB}} | \mathcal{N})$. As mentioned earlier, we construct $p(R_{\mathrm{GRB}} | \mathcal{N})$ separately for sGRBs and lGRBs in each cluster, and add the rate posteriors sample by sample, to calculate the total rate for that (BNS or NSBH) cluster. We plot the BNS (NSBH) rate posterior in the left (right) panel of Figure~\ref{fig:rates}. For comparison, we also plot the BNS (NSBH) GW rate posterior, $p(R_{\mathrm{GW}} | d_{\mathrm{GW}})$, as estimated in Appendix C of \citet{KAGRA:2021duu}\footnote{Note that Appendix C of \citet{KAGRA:2021duu} only provides the upper and lower limits on the GW rate posteriors, and not the full posteriors. We plot these for completeness and also evaluate the medians.}.

We find that the posteriors on the rate of GRB-Bright BNSs (NSBHs), acquired from {\it Fermi}/GBM data, yield a median and $90\%$ confidence interval of $50^{+199}_{-44} \mathrm{Gpc}^{-3}\mathrm{yr}^{-1}$ ($5^{+18}_{-4} \mathrm{Gpc}^{-3}\mathrm{yr}^{-1}$). The corresponding {\it Swift}/BAT rate estimates are $13^{+45}_{-11} \mathrm{Gpc}^{-3}\mathrm{yr}^{-1}$ ($1^{+4}_{-1} \mathrm{Gpc}^{-3}\mathrm{yr}^{-1}$). The GRB-Bright BNS rate is consistent with the LVK's rate estimates given in Appendix C of \citet{KAGRA:2021duu}: $163^{+300}_{-134} \mathrm{Gpc}^{-3}\mathrm{yr}^{-1}$. On the other hand, the GRB-Bright NSBH rate is consistent with the expectation that not all NSBHs produce EM-counterparts, with the GW NSBH rate being estimated to be  ($68^{+120}_{-53} \mathrm{Gpc}^{-3}\mathrm{yr}^{-1}$).

Assuming all BNSs produce GRBs, we construct the distribution of beaming fractions (Figure~\ref{fig:fraction}, left panel) and angles (Figure~\ref{fig:fraction}, right panel), as described in the previous section. We constrain the beaming angle to $\theta_j \in [\theta_j^{\mathrm{min}} = 6.4{\degree}^{+13.4\degree}_{-5.0\degree}, \theta_j^{\mathrm{max}} = 12.2{\degree}^{+21.3\degree}_{-8.2\degree}]$ ($\theta_j \in [\theta_j^{\mathrm{min}} = 3.3{\degree}^{+6.4\degree}_{-2.5\degree}, \theta_j^{\mathrm{max}} = 6.2{\degree}^{+10.0\degree}_{-4.0\degree}]$), at $90\%$ confidence, with {\it Fermi}/GBM ({\it Swift}/BAT) data. Across both data sets, the beaming angle is constrained to $\theta_j \in [0.8\degree, 33.5\degree]$. The tighter constraints on the GRB rate and beaming angle estimates from {\it Swift}, in comparison to {\it Fermi}, are a direct consequence of the larger uncertainty in the fraction of events in the BNS/NSBH cluster that can be confidently ascertained to have a BNS/NSBH provenance in {\it Fermi} data.

While the assumption that all BNS mergers produce GRBs has been assumed in the literature (see, e.g., \cite{williams2018constraints}), in reality, this will depend on whether the merger remnant is a {\it black hole}, {\it hypermassive neutron star} which undergoes a delayed collapse to a  black hole or a long-lasting {\it supramassive neutron star}. Recently, for instance, \cite{Bamber:2024kfb} argued from numerical simulations that a supramassive NS with an accretion disk may not necessarily produce a jetted emission unlike a BH or a hypermassive NS. This would indicate that the mass, spin and the equation of state of the binary constituents will decide the nature of the remnant and the emergence of a jet from it. Therefore an unknown fraction of binary neutron star mergers may not produce GRBs.
If true, our constraints on the beaming fraction would be systematically underestimated \footnote{It is interesting to constrain the fraction of GRB-Bright BNSs, employing the same method used for estimating this fraction for NSBHs. Assuming, as before, the distribution of beaming fractions to be uniformly distributed in $f_b \in [0.01, 0.10]$, we constrain the fraction of GRB-Bright BNSs from {\it Swift} data to be $f_B \in [1\%, 75.6\%]$. This is consistent with \citep{Salafia2022}, that find the fraction to be $\gtrsim 1/5$.}

We also construct the distribution on the fraction $f_B$ of GRB-Bright NSBHs, and constrain it, at $90\%$ confidence, to $f_{B} \in [1.3\%, 63\%]$ ($f_B \in [0.4\%, 15\%]$), with {\it Fermi} ({\it Swift}) data. These are consistent with other estimates of $f_B$ in the literature, such as NSBH population driven constraints of \citep{Biscoveanu2023}, and population synthesis driven constraints of \citep{Broekgaarden2021}.

Past works \citep{williams2018constraints, Biscoveanu:2019bpy, Mogushi:2018ufy, farah2020counting, Sarin:2022cmu, hayes2023unpacking} that place constraints, including prospective ones with future observations, assume that all sGRBs are associated with CBCs containing at least one NS. On the other hand, our work provides a potentially powerful means to constrain the fraction of GRB-Bright NSBHs, and the beaming angle of BNS-associated GRBs, using results from the ML-based classification scheme of \citet{Dimple:2023wvs, Dimple_2024}. The main advantages of our method are that: 1) it does not require spatial and temporal coincidence of individual GW and GRB events, thus mitigating the lack of joint GW-GRB detections; 2) it does not assume that all sGRBs are associated with BNSs/NSBHs.

Though modest, the constraints on the fraction of GRB-Bright NSBHs will improve with increased NSBH associated GW detections from the LVK, as well as GRBs whose KN/SN provenance is observed, to reduce the uncertainty in the labelling of the events in the NSBH cluster. Such constraints will have significant implications on our understanding of GRB production in NSBH systems, as well as more profound implications on NS EOS constraints. 

The constraints on the beaming angle $\theta_j$ are consistent with corresponding estimates reported in the literature. Observations suggest that sGRB beaming angles lie within $\theta_j \in [1\degree, 30\degree]$ \citep{soderberg2006afterglow, burrows2006jet, fong2012jet,  coward2012swift}. This is largely corroborated by simulations \citep{rosswog2002jets, janka2006off, rezzolla2011missing}. As with the constraints on $f_B$ we expect the constraints on $\theta_j$ to improve with additional detections of KNe/SNe associated with GRBs in the BNS cluster. 

There are two important caveats to our work. The first is the assumption that all BNS/NSBH-associated GRBs have been identified and placed in the appropriate clusters by the ML algorithm. This, however, has not been confirmed. Thus, while false-positives have been accounted for, false negatives\footnote{BNS/NSBH-associated GRBs that have not been identified by the ML algorithm.} have not. Moreover, recent numerical simulations (e.g. \citep{Gottlieb2023}) suggest that distinguishing properties of GRB light curves are not necessarily driven by their progenitors, but certain intrinsic properties of the postmerger that could be produced by both BNSs and NSBHs. Consequently, the ML algorithm, which is currently not trained on these simulations, could be clustering the data in a manner that does not accurately segragate BNSs from NSBHs. These could potentially contribute to systematics in the BNS/NSBH-associated GRB rate estimation. We expect that more sophisticated ML methods employed to classify GRB light curves and additional joint-detections of BNS/NSBH-associated GWs and GRBs will help mitigate said systematics.

The second is that our analysis neglects the redshift evolution of the GRB rate. This was done to be consistent with the method used to estimate the GW merger rate for BNSs/NSBHs, although redshift-evolving GRB rate estimates also have significant uncertainties arising from a largely unconstrained delay-time distribution of CBCs. As mentioned earlier, currently, the BNS/NSBH sensitivity range of LVK detectors does not probe large enough redshifts to determine a corresponding redshift evolution. However, with improved sensitivity in future observing scenarios \citep[e.g: XG detector network, ][]{punturo2010, Reitze:2019iox}, we expect that our ML-based method to constrain the rate of BNS/NSBH-associated GRBs and correspondingly the beaming angle/GRB-bright fraction, updated to incorporate a redshift evolution, will have significant astrophysical implications. 

%
\section*{Acknowledgements}
We thank Michael Coughlin for their feedback on the manuscript, and Anarya Ray for help with the GW rate posteriors. We thank the Indian Institute of Technology (IIT), Bombay, for organizing the ``Transients 2024" workshop during which this project was initiated. SJK acknowledges support from the Science and Engineering Research Board (SERB) Grant SRG/2023/000419. 
D and KGA acknowledge support from the Department of Science and Technology (DST) and the SERB of India via the Swarnajayanti Fellowship Grant DST/SJF/PSA-01/2017-18. They also acknowledge support from a grant from the Infosys Foundation.
RL, KM and KGA acknowledge the support from BRICS grant DST/ICD/BRICS/Call-5/CoNMuTraMO/2023 (G) funded by the DST, India. 

This research has made use of data or software obtained from the Gravitational Wave Open Science Center (gwosc.org), a service of the LIGO Scientific Collaboration, the Virgo Collaboration, and KAGRA. This material is based upon work supported by NSF's LIGO Laboratory, which is a major facility fully funded by the National Science Foundation, as well as the Science and Technology Facilities Council (STFC) of the United Kingdom, the Max-Planck-Society (MPS), and the State of Niedersachsen/Germany for support of the construction of Advanced LIGO and construction and operation of the GEO600 detector. Additional support for Advanced LIGO was provided by the Australian Research Council. Virgo is funded, through the European Gravitational Observatory (EGO), by the French Centre National de Recherche Scientifique (CNRS), the Italian Istituto Nazionale di Fisica Nucleare (INFN) and the Dutch Nikhef, with contributions by institutions from Belgium, Germany, Greece, Hungary, Ireland, Japan, Monaco, Poland, Portugal, Spain. KAGRA is supported by Ministry of Education, Culture, Sports, Science and Technology (MEXT), Japan Society for the Promotion of Science (JSPS) in Japan; National Research Foundation (NRF) and Ministry of Science and ICT (MSIT) in Korea; Academia Sinica (AS) and National Science and Technology Council (NSTC) in Taiwan.
\vspace{5mm}

\textit{Software}: \texttt{NumPy} \citep{vanderWalt:2011bqk}, \texttt{SciPy} \citep{Virtanen:2019joe}, \texttt{astropy} \citep{2013A&A...558A..33A, 2018AJ....156..123A}, \texttt{Matplotlib} \citep{Hunter:2007}, \texttt{jupyter} \citep{jupyter}.

\bibliography{references}

\end{document}